\documentclass[amsmath,amssymb,
               aps,
               prc,
               twocolumn,
               superscriptaddress,
               showpacs,
               nofootinbib
]{revtex4-2}
\usepackage[utf8]{inputenc}
\usepackage{graphicx}
\usepackage{color}  
\usepackage{amsmath}
\usepackage{dcolumn}
\usepackage[english]{babel}
\usepackage[colorlinks=true,linkcolor=blue,citecolor=blue, urlcolor=blue]{hyperref}
\usepackage[dvipsnames]{xcolor}
\usepackage{multirow}
\usepackage{verbatim}
\usepackage{lipsum}
\usepackage{xspace}
\usepackage{braket}

\newcommand{\emax}{\text{e}_\text{1max}\xspace}
\newcommand{\etmax}{\text{e}_\text{3max}\xspace}

\newcommand{\tndim}{\tilde{n}_\text{dim}}
\newcommand{\ndim}{n_\text{dim}}

\newcommand{\ai}{\emph{ab initio}}

\newcommand{\magic}{1.8/2.0 (EM)\xspace}
\newcommand{\go}{{\Delta}{\rm NNLO}_{\rm GO}(394)\xspace}

\begin{document}

\title{Deformed self-consistent Green's function method for atomic nuclei\\ at second and third order in the algebraic diagrammatic construction}

\author{Alberto~Scalesi}
    \email{alberto.scalesi@chalmers.se}
    \affiliation{Department of Physics and Astronomy, Chalmers University of Technology, SE-412 96 G\"oteborg, Sweden}
\author{Thomas~Duguet}
	\affiliation{Universit\'e Paris-Saclay, CEA, IRFU, 91191 Gif-sur-Yvette, France}
\author{Vittorio~Som\`a}
	\affiliation{Universit\'e Paris-Saclay, CEA, IRFU, 91191 Gif-sur-Yvette, France}

\date{\today}

\begin{abstract}
\begin{center}
\begin{minipage}{0.85\textwidth}
\begin{description}
\item[Background] The description of atomic nuclei from first principles constitutes one of the main aims of nuclear theory. Exact solutions of the $A$-body Schr\"odinger equation based on realistic inter-nucleon interactions are currently limited to light nuclei due to the exponential computational scaling with system size. Over the last fifteen years, polynomial-scaling expansion methods have however extended \textit{ab initio} calculations at sub-percent accuracy to all medium-mass plus a few (spherical) closed-shell heavy nuclei. 
\item[Purpose] The largest portion of the Segr\`e chart is constituted of (deformed) doubly open-shell heavy and superheavy systems. While these are currently largely out of reach of \textit{ab initio}  nuclear structure theory, the objective is to develop formal many-body methods and associated numerical codes that make their description from first principles possible.
\item[Methods] The self-consistent Green's function formalism is extended to doubly open-shell nuclei by allowing the one-body propagator to spontaneously break SU(2) rotational symmetry.
The resulting deformed scheme, based on the algebraic diagrammatic construction truncation at first, second, and third order, is implemented in a newly developed many-body suite.
The numerical strategies required to handle the $M$-scheme working basis, whose dimension raises the computational cost by orders of magnitude with respect to a symmetry-restricted implementation, are discussed in detail. The potential and rationale of deformed self-consistent Green's function calculations of doubly open-shell nuclei are illustrated through a study of $^{28}$Si based on the \magic Hamiltonian.
\item[Results] First, the impact of the three-nucleon interaction, and of its approximate treatment via rank-reduction techniques, on the topology of the deformed Hartree-Fock  total energy curve produced as a function of axial quadrupole deformation is scrutinized. Second, the topology of the {\it correlated} total energy curve produced via constrained self-consistent Green's function calculations is shown to differ appreciably from the underlying deformed Hartree-Fock one. Third, physical solutions appearing as minima of the correlated total energy curve are shown to also be reachable via unconstrained calculations starting from any of the states belonging to the deformed Hartree-Fock total energy curve. This demonstrates the genuinely self-consistent character of the method. Fourth, the $^{28}$Si ground-state binding at third order in the many-body expansion associated with the oblate solution is extrapolated to the infinite basis-size limit to reproduce experiment within $2.1\%$, while the prolate solution appears to be consistent with the observed $0^+_3$ shape isomer.
\item[Conclusions] Calculations based on deformed self-consistent Green's function theory provide a new path towards doubly open-shell nuclei. The way such calculations access nuclear ground states and shape isomers is illustrated using $^{28}$Si as a characteristic example. The present developments open the way to an accurate   \textit{ab initio}  description of all (very) heavy nuclei in the near future.
\end{description}
\end{minipage}
\end{center}
\end{abstract}

\maketitle

\section{Introduction}
\label{sec:intro}

Describing atomic nuclei starting from a Hamiltonian rooted in quantum chromodynamics via  effective field theories constitutes the central goal of \ai{} nuclear structure theory.
The systematic construction of two- and three-nucleon interactions within chiral effective field theory, combined with renormalization-group techniques softening their short-range content~\cite{Hebeler21}, aims at providing the necessary input for such an endeavor.
On the many-body side, a decisive step has been the development of correlation-expansion methods whose computational cost scales polynomially rather than exponentially with the size of the system~\cite{Hagen14,Hergert16,Soma20b,Tichai20b}.
Together, these advances have extended the reach of \ai{} calculations from the lightest systems to the medium-mass region and, more recently, to the heavy-mass sector~\cite{Bonaiti2025heavy,Demol26,Vernik2026}.
 
This progress has long been restricted to closed-shell nuclei.
The reason is that correlation-expansion methods typically expand the exact solution on top of a symmetry-conserving reference state that must be non-degenerate with respect to its elementary excitations for that expansion to be controlled.  This is however not the case in open-shell nuclei where the valence shell in the single-particle spectrum associated with such a reference state is only partially occupied such that the expansion breaks down in such systems.
The most efficient way out beyond the lightest nuclei consists in allowing the reference state to spontaneously break symmetries of the Hamiltonian, thereby lifting the problematic degeneracy at the price of working with a state that no longer carries good symmetry quantum numbers~\cite{Duguet15a,Duguet17a}.
Breaking U(1) particle-number symmetry captures strong static pairing correlations and grants access to singly open-shell nuclei, whereas breaking SU(2) rotational invariance captures strong static quadrupole correlations driving the intrinsic nuclear deformation and is thus mandatory to address doubly open-shell systems~\cite{Scalesi:2024nao}.
This approach, imported from energy-density-functional theory, has been implemented over the past decade in several expansion frameworks, from many-body perturbation theory~\cite{Tichai18a,Frosini21,Frosini22c} to coupled-cluster~\cite{Signoracci15,Novario20,Hagen22,Tichai23} and in-medium similarity renormalization group~\cite{Sun25,Demol26} approaches. As a matter of fact, the strategy to expand solutions of the many-body Schr\"odinger equation with respect to a symmetry-breaking reference state was initiated within self-consistent Green's function (SCGF) theory to address singly open-shell nuclei via the U(1)-breaking Gorkov extended scheme~\cite{Soma11,Soma14a,Soma20a}. While the SU(2)-breaking counterpart has been missing so far, it is the purpose of the present work to report on its implementation and first applications, demonstrating how calculations based on the new associated numerical suite open a new path towards the {\it ab initio} description of all nuclei on the Segr\`e chart. 

The SCGF formalism is thus generalized to propagators breaking rotational symmetry, a scheme referred to as deformed SCGF (dSCGF). Doing so brings key novel capabilities compared to similar calculations based on other expansion methods~\cite{Frosini21,Novario20,Frosini22c,Hagen22}.
First, the one-body propagator provides direct access to the one-nucleon addition and removal spectroscopy associated to  all final states in $A\pm1$ neighbors and constitutes a natural starting point for the \ai{} construction of optical potentials in doubly open-shell nuclei.
Second, the SCGF scheme is genuinely self-consistent: the dressed propagator is determined by a fixed-point equation, so that the final solution is fully independent of the reference state used to initialize the calculation\footnote{As explained above, the reference state must however offer an appropriate starting point by breaking rotational symmetry.}.
This feature turns out to be particularly valuable in doubly open-shell nuclei, where several deformed mean-field minima compete and where the connection between mean-field and correlated solutions is not of a one-to-one kind as illustrated below.
Third, the associated $\Phi$-derivable character~\cite{Soma11} endows the second-order truncation with a variational character that can be exploited by scanning the intrinsic deformation through the notion of total energy curve to access both the ground-state as well as shape isomers. Fourth, the working dSCGF equations are amenable to a very efficient numerical implementation, allowing one to address the runtime and memory challenges associated with the very large working bases needed to converge calculations of (very) heavy nuclei.
 
In this work, the dynamical self-energy is expanded within the algebraic diagrammatic construction scheme~\cite{Schirmer83,SchirmerBook,Schirmer89} such that dSCGF($n=1,2,3$) calculations are performed, the lowest order ($n=1$) corresponding to deformed Hartree-Fock (dHF) calculations. The method is implemented in \texttt{FoxTrot}, a newly developed many-body suite designed from the outset for symmetry-unrestricted calculations.
Because one-body propagators are no longer diagonal in the total angular momentum and its projection, the working one-body basis must be handled in full, leading to a dimension that typically exceeds by one order of magnitude the effective dimension at play in a symmetry-restricted implementation~\cite{Frosini24}.
Given the polynomial scaling of dSCGF($2$) and dSCGF($3$) approximation schemes, this amounts to an increase of the computational cost by several orders of magnitude, which makes the numerical implementation as much a crucial part of the present development as the formalism itself.
 
The formalism and its numerical realization are illustrated here through a detailed study of $^{28}$Si. This nucleus constitutes an ideal testing ground: it is light enough to allow for dSCGF($3$) calculations in sizable bases, yet it displays a genuine shape-coexistence pattern, with an oblate ground state and a $J^{\pi}=0^+$ prolate shape isomer at $E_\text{x}=6.69$\,MeV~\cite{Endt:1973div,kanada03,Frycz24,Taniguchi:2026hnj}.
It thus allows one to establish, in a controlled setting, the procedures needed to generate, identify and follow multiple correlated solutions, to assess their convergence with respect to basis-size parameters, and to quantify the impact of the rank-reduction treatment of three-nucleon interactions~\cite{Frosini21}. Note that the accuracy of the dSCGF method has already been partly assessed in Refs.~\cite{scalesi_thesis,Fu26}.
 
The article is organized as follows.
Section~\ref{sec:dscgf} presents the dSCGF formalism, the symmetry-breaking scheme, constrained calculations delivering correlated total energy curves, the self-consistency algorithm and the calculation of the total energy.
Section~\ref{sec:numerics} discusses the numerical implementation in view of the computational challenge associated with the breaking of rotational symmetry.
Section~\ref{28SiAnalysis} is devoted to the application to $^{28}$Si.
Conclusions and perspectives are collected in Sec.~\ref{sec:conclusions}.

\section{Theory}
\label{sec:dscgf}

\subsection{SCGF formalism}

The self-consistent Green's function approach~\cite{DickhoffVanNeck} tackles the $A$-body problem via a correlation expansion truncated to produce systematically improvable solutions.
In this framework, the key quantity is the dressed one-body propagator $g$, a two-point correlation function encapsulating the effects of the intricate particle interactions inside the many-body system.
The one-body propagator is obtained by solving Dyson's equation
\begin{equation}
g_{\alpha\beta}(\omega) = g^{(0)}_{\alpha\beta}(\omega) + \sum_{\gamma\delta} g^{(0)}_{\alpha\gamma}(\omega)\, \Sigma^{\star}_{\gamma\delta}(\omega)\, g_{\delta\beta}(\omega) \; , \label{dyson}
\end{equation}
where $g^{(0)}$ denotes some initial ansatz for $g$, typically obtained from an HF calculation. 
Greek indices label a complete orthonormal one-body basis $\{ | \alpha \rangle \}$, with associated creation (annihilation) operators $\{ a^\dagger_\alpha \}$ ($\{ a_\alpha \}$) whereas $\omega$ represents an energy variable. 

In Eq.~\eqref{dyson}, the irreducible self-energy $\Sigma^{\star}$, encoding all interactions contributing to the dressed one-body Green's function, can be decomposed according to 
\begin{equation}
\Sigma^{\star}_{\alpha\beta}(\omega) \equiv \Sigma^{(\infty)}_{\alpha\beta} + \tilde{\Sigma}_{\alpha\beta}(\omega) \; , 
\end{equation}
where $\Sigma^{(\infty)}$ represents the $\omega$-independent static part and $\tilde{\Sigma}(\omega)$ accounts for dynamical correlations. The latter object is expanded and truncated to render the many-body computation tractable. While the expansion of the dynamical self-energy can rely on simple perturbation theory, such truncations typically do not satisfy the analytic form of the exact  $\tilde{\Sigma}$ reading as
\begin{align}
\tilde{\Sigma}_{\alpha\beta}(\omega) \equiv&
\sum_{rr'} M^\dagger_{\alpha r} \left [ \frac{1}{\omega - (E^>+C) + i \eta} \right ]_{rr'} M_{r' \beta} 
\nonumber
\\ +&
\sum_{ss'} N^\dagger_{\alpha s} \left [ \frac{1}{\omega - (E^<+D) - i \eta} \right ]_{ss'} N_{s' \beta} \; .
\label{eq_sigma_Lehmann}
\end{align}
In Eq.~\eqref{eq_sigma_Lehmann}, matrices $(M,N)$ couple the single-particle motion of the nucleon to intermediate multiparticle-multihole configurations (labelled by composite indices $r$ and $s$), whose energies (`bare' and resulting from the interference between such configurations) are encoded into matrices $(E^>,E^<)$ and $(C,D)$, respectively. 
The presently employed expansion scheme, the algebraic diagrammatic construction (ADC)~\cite{Schirmer83,SchirmerBook}, imposes instead that the truncated self-energy displays the correct analytic structure by directly devising an expansion of the building blocks in Eq.~\eqref{eq_sigma_Lehmann}.
At order ADC($n$), matrices $M$, $N$, $C$ and $D$ are determined by demanding that all perturbation-theory contributions up to order $n$ are included.
Combining this requirement with the structure of Eq.~\eqref{eq_sigma_Lehmann}, specific sets of perturbative contributions are resummed to all orders into the self energy for ADC($n \ge 3$).

The preservation of the analytical form of the exact self-energy at each order $n$ guarantees key properties. First, causality is not violated. Second, this form allows the derivation of an energy-independent auxiliary eigenvalue problem that significantly simplifies the numerical solution of Dyson's equation (see later).
Last but not least, the resulting solutions are free from pathological behaviors~\cite{Hirata24}.

In this work, ADC($n=1,2,3$) truncations are considered and define  so-called SCGF($n=1,2,3$) calculations.
While the first order, SCGF(1), corresponds to the standard HF approximation, 
SCGF($2$) incorporates lowest-order dynamical correlations associated with two-particle/one-hole and two-hole/one-particle intermediate state configurations. 
The ADC($3$) truncation further introduces couplings between such configurations and, as a result, includes infinite-order resummations of both particle-particle/hole-hole and particle-hole ladders.
The ADC expansion delivers a systematic approximation scheme  of choice in quantum chemistry~\cite{Banerjee23,Nanni23}, where it has been shown that SCGF($2$) and SCGF($3$) results display respectively percent and sub-percent accuracy~\cite{Barbieri12,Dempwolff22}. 
Applications in closed-shell systems have been shown to deliver a similar level of accuracy for nuclear structure applications~\cite{Duguet17b,Soma20a}.

Once ADC($n$) matrices $M$, $N$, $C$ and $D$ have been constructed, the analytic form~\eqref{eq_sigma_Lehmann} of the self-energy allows one to recast Dyson's equation into an energy-independent eigenvalue problem consisting in the diagonalization of the effective Hamiltonian~\cite{Schirmer89}
\begin{equation}
H^{\text{ADC}} \equiv
\begin{pmatrix}
H^{(1)} + \Sigma^{(\infty)} & M^{\dagger} & N^{\dagger} \\[2pt]
M & E^{>} + C & 0 \\[2pt]
N & 0 & E^{<} + D
\end{pmatrix}
\, ,
\label{eq_ADC_matrix}
\end{equation}
where $H^{(1)}$ is the one-body part of the nuclear Hamiltonian that is presently considered to include one-nucleon (1N), two-nucleon (2N) and three-nucleon (3N) operators
\begin{equation}
H \equiv \sum_{k=1}^3 H^{(k)} \, .
\end{equation}
Eigenvalues
\begin{subequations}
\label{eq_poles}
\begin{eqnarray}
\varepsilon^+_{k} &\equiv& E^{A+1}_k - E^A_0  \; , \\
\varepsilon^-_{l} &\equiv& E^A_0 - E^{A-1}_l \; ,
\end{eqnarray}
\end{subequations}
and eigenvectors
\begin{subequations}
\label{eq_amplitudes}
\begin{eqnarray}
\mathcal{X}^{k}_\alpha &\equiv& \langle\Psi^{A+1}_k\vert  a^\dagger_\alpha  \vert\Psi^A_0 \rangle  \; , \\
\mathcal{Y}^{l}_\alpha &\equiv& \langle\Psi^{A-1}_l\vert  a_\alpha  \vert\Psi^A_0\rangle \; ,
\end{eqnarray}
\end{subequations}
of $H^\text{ADC}$ constitute the building blocks of the K\"all\'en-Lehmann spectral representation~\cite{Kallen52,Lehmann54} of the one-body Green's function according to
\begin{align}
 g_{\alpha\beta}(\omega)
  ={}& \sum_k \frac{(\mathcal{X}^{k}_\alpha)^*  \mathcal{X}^{k}_\beta}{\omega  - \varepsilon^+_{k} + \textrm{i}\eta} 
        ~+~ \sum_l \frac{\mathcal{Y}^{l}_\alpha  (\mathcal{Y}^{l}_\beta)^*}{\omega  - \varepsilon^-_{l} - \textrm{i}\eta}  \; .
\label{eq_Lehmann}
\end{align}

\subsection{Breaking rotational invariance}

The present work extends SCGF to doubly open-shell systems by allowing the breaking of SU(2) rotational invariance associated with the conservation of total angular-momentum. Indeed, this allows one to capture, already at the ADC(1) level, strong static angular (e.g.~quadrupole) correlations driving the phenomenology associated with intrinsic nuclear deformation(s).  The  \texttt{FoxTrot} code discussed below allows the breaking of any spatial symmetry (parity, angular momentum, or its projection along the quantization axis) and isospin, i.e.\ the most general parity-breaking triaxial dHF propagator can be used as an input for the ADC$(2,3)$ routines. However, present applications (and the following discussion) are restricted to parity-conserving axial deformations.

Symmetries broken in the first place must eventually be restored~\cite{Duguet15a,Duguet17a} to obtain quantities (and eventually observables) characterized by good symmetry quantum numbers. While dSCGF($n$) solutions do carry good angular momentum in the exact limit  ($n\rightarrow \infty$), those obtained at low truncation order ($n=1,2,3\ldots$) do not. Consequently, restoring good angular momentum requires an explicit projection technique in practical calculations. While this has been formulated within many-body perturbation theory and the coupled cluster method~\cite{Duguet15a,Hagen22}, it remains to be done for SCGF theory. The impact of angular momentum restoration on bulk properties presented in this paper is expected to be negligible compared to  other uncertainties characterizing current ab initio many-body calculations~\cite{Hagen22}.

\subsection{Deformation and constrained calculations}

The breaking of rotational symmetry of any targeted $J=0$ many-body state (or one-body propagator, in the present context) is presently quantified by computing the intrinsic axial deformation of even multipolarity $\lambda$ based on the one-body axial multipole moment operator defined as
\begin{equation}
Q_{\lambda0} \equiv \sum_{i=1}^A r_i^{\lambda} \, Y_{\lambda0}(\theta_i,\varphi_i) \equiv \sum_{\alpha\beta} q^{(\lambda 0)}_{\alpha\beta} \; a^\dagger_\alpha a_\beta \; .
\end{equation}
The symmetry breaking is signaled by non-zero expectation values
\begin{equation}
\langle Q_{\lambda0} \rangle  
\equiv \sum_{\alpha\beta} \int_{C \uparrow} \frac{d\omega}{2\pi i} \, q^{(\lambda 0)}_{\alpha\beta}~g_{\beta\alpha}(\omega)
\equiv\sum_{\alpha\beta}  q^{(\lambda 0)}_{\alpha\beta}~\rho_{\beta\alpha} \; ,
\label{eq_Q_expectation}
\end{equation}
where $\rho$ denotes the one-body density matrix associated with the many-body state under consideration.
The axial quadrupole ($\lambda=2$) deformation of present interest is eventually quantified via the dimensionless parameter
\begin{equation}
\beta_2 \equiv \frac{4\pi}{3 A R_0^2} \langle Q_{20} \rangle \; , 
\label{eq_beta}
\end{equation}
with $R_0 \equiv 1.2\, A^{1/3}\ \text{fm}$. In practical calculations, the notation $\beta^{\text{dSCGF}(n)}_2$ will be used whenever the expectation value on the right-hand side of Eq.~\eqref{eq_beta} is computed at the dSCGF($n$) level.

The first step of a dSCGF($2$) or dSCGF($3$) computation consists in a dHF=dSCGF(1) calculation.
This can be either a single variational (unrestricted) dHF run, or a series of dHF calculations constrained to different axial quadrupole deformations.
Constrained calculations are performed by replacing the Hamiltonian with the Routhian
\begin{equation}
H' \equiv H - \mu_{q_{20}} Q_{20} \; ,
\label{eq_routhian}
\end{equation}
where $\mu_{q_{20}}$ denotes a Lagrange multiplier set to impose the required $\beta_2^\text{dHF}$ value.
The latter procedure delivers a dHF total energy curve (TEC) as a function of $\beta_2^\text{dHF}$, similarly to what is traditionally done in energy density functional (EDF) calculations~\cite{Bender03}.

The procedure provides a (set of) starting point(s) for the subsequent dSCGF($2,3$) evaluation, for which three possibilities exist, namely
\begin{enumerate}
\item The dSCGF($2,3$) calculation starts from the variational dHF propagator, i.e.\ the global minimum of the dHF TEC;
\item A series of unconstrained dSCGF($2,3$) calculations are performed on top of each point in the dHF TEC;
\item A set of constrained dSCGF($2,3$) calculations\footnote{Constrained dSCGF($n>1$) calculations proceed analogously to dHF, i.e.\ by replacing H with the Routhian~\eqref{eq_routhian}. This results in the substitution
$H^{(1)} \rightarrow H^{(1)} - \mu_{q_{20}} Q_{20}$ into Eq.~\eqref{eq_ADC_matrix}.} are performed\footnote{In principle any dHF propagator can be used as a starting point for this step. In practice, starting from a dHF solution constrained to a $\beta^{\text{dHF}}_2$ value close to the targeted $\beta_2^\text{dSCGF($n$)}$ one results in a much faster convergence.}. 
\end{enumerate}

Option 3 delivers a {\it correlated} TEC at the dSCGF($2,3$) level, analogous to what is obtained at the uncorrelated dHF level.
Option 2 yields a finite set of solutions corresponding to the various minima of the dSCGF($2,3$) TEC. Notably, the solution associated with a given minimum is typically reached several times when starting from the various points along the dHF TEC.
In option 1, the dSCGF($2,3$) calculation converges to one of the minima of the corresponding TEC.  While the latter is usually the global minimum, it is possible that a local minimum is reached instead. These different procedures are illustrated in Sec.~\ref{28SiAnalysis}, dedicated to $^{28}$Si.

\subsection{Self-consistency}

For any of the options discussed above, Dyson's equation is solved self-consistently according to the following scheme:
\begin{enumerate}
\item A dHF reference state is generated;
\item The reference-state propagator is used to construct the second- and, if required, third-order self-energy diagrams, i.e.\ to compute the ADC($2,3$) matrix~\eqref{eq_ADC_matrix};
\item Owing to its large dimensions, the ADC($2,3$) matrix cannot be diagonalized directly. Hence, it is first compressed by means of a Krylov projection, which preserves the pole structure in the vicinity of the Fermi surface, to which the spectral strength is most sensitive~\cite{Soma14a};
\item The ADC($2,3$) matrix is then diagonalized iteratively through a set of inner loops where  the static part $H^{(1)} + \Sigma^{(\infty)}$ is regenerated at the end of each iteration. Upon convergence, this defines the dSCGF$_0(2,3)$ approximation;
\item Full self-consistency, i.e., including iterative generation of dynamical contributions, is achieved by generating an optimized reference state (OpRS)~\cite{Raimondi18} as an intermediate that compresses the information contained in the dSCGF($2,3$) propagator into an effective mean-field-like propagator. The latter is used as a new input to step 2 and defines a second outer loop. Denoting by $k$ the number of outer loops thus generated, i.e.\ the number of times the second- and third-order terms are recomputed based on the intermediate OpRS, the dSCGF$_k(2,3)$ sequence converges to a fixed point that is independent of the OpRS and defines the fully self-consistent dSCGF($2,3$) solution.
\end{enumerate}

As the above algorithm stipulates, a dSCGF($2,3$) calculation is initialized from a preliminary dHF solution. The method is nevertheless independent of that choice: since first-order contributions are recomputed at each inner iteration from the dressed propagator, any reference state can in principle be used to reach the self-consistent solution. 
A poor starting point affects only the numerical behavior of the outer loops, which may converge slowly or, in extreme cases, become unstable.

\begin{figure}
    \includegraphics[scale=0.8]{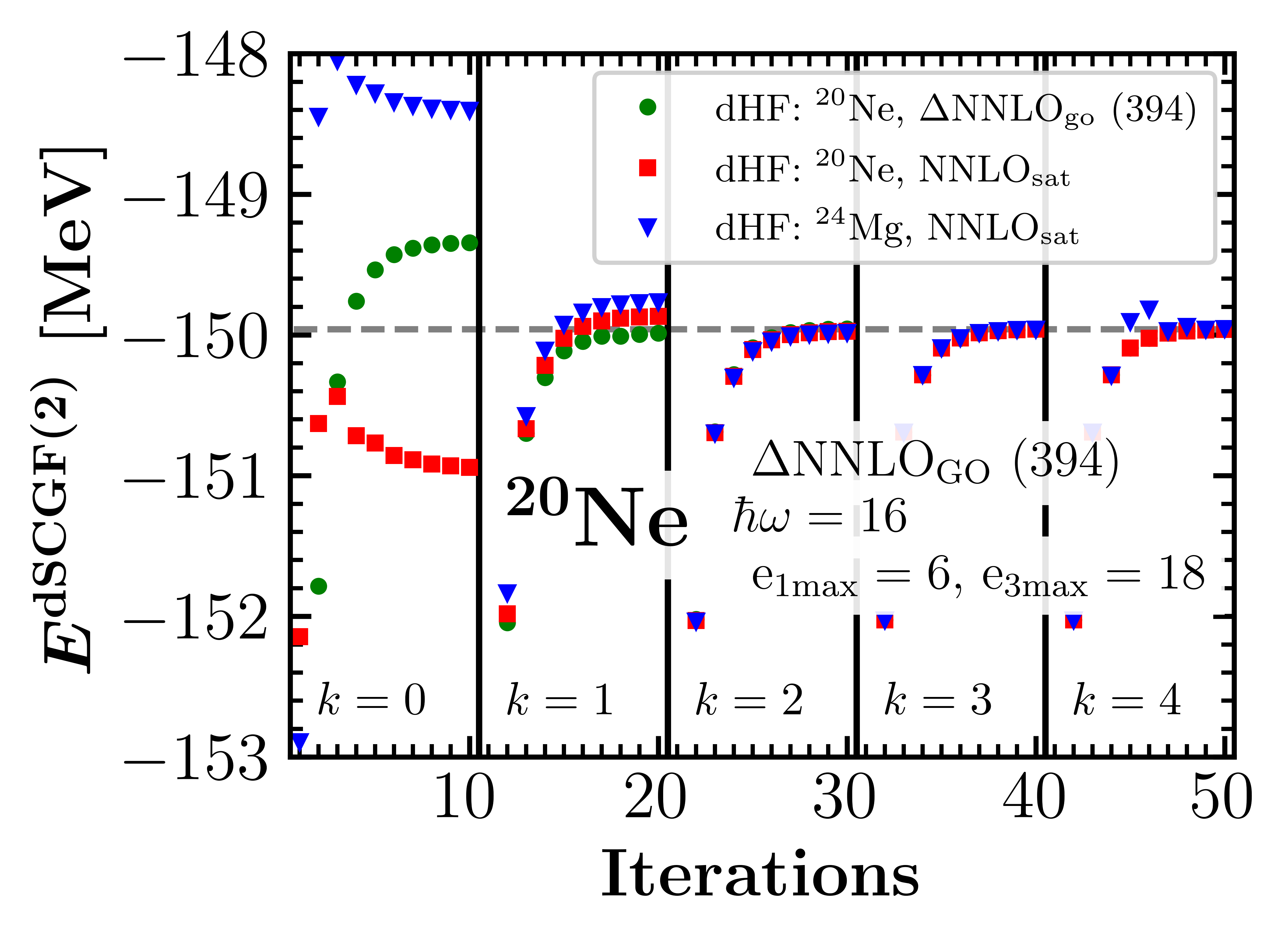}
	\caption{Numerical demonstration of self-consistency in dSCGF($2$) calculations of $^{20}$Ne based on the $\Delta$NNLO$_{\rm GO}$(394) Hamiltonian~\cite{Jiang20} and $(\hbar\omega, \emax, \etmax) = (16\,\text{MeV},6,18)$. The dSCGF$_k(2)$ ($k=0-4$) binding energy is displayed for three different dHF reference states (obtained for the nucleus and Hamiltonian indicated in the legend). Values of $k$ designate the number of outer iterations of the self-consistent process whereas the x-axis denotes the cumulative number of inner iterations inside the outer iterations.}
	\label{fig_SC}
\end{figure}

This independence is illustrated numerically in Fig.~\ref{fig_SC} displaying the convergence pattern of dSCGF$_k(2)$ calculations of $^{20}$Ne based on the $\Delta$NNLO$_{\rm GO}$(394) Hamiltonian~\cite{Jiang20} as a function of $k$. Calculations are repeated starting from three different dHF reference states, obtained for a different nucleus ($^{24}$Mg) and/or a different Hamiltonian (NNLO$_{\rm sat}$~\cite{Ekstrom15}). Starting from the reference state computed from the same nucleus and Hamiltonian yields the fastest convergence, i.e.\ the $k=0$ result already lies closest to the fully self-consistent solution (dashed line) and the number of iterations needed to reach self-consistency is minimal. Still, the remarkable fact is that all sequences converge to the same dSCGF($2$) solution, irrespective of the starting state.

\subsection{Total energy}

Once the correlated one-body propagator $g$ has been obtained, one-body observables can be computed straightforwardly from it. Additionally, the total energy of the $A$-body system can  be computed via the Galitskii-Migdal-Koltun (GMK) sum rule~\cite{Galitskii58,Koltun72} corrected for the presence of three-body forces~\cite{Carbone13} 
\begin{equation}
E^A = \sum_{\alpha\beta} \frac{1}{2} 
\int_{C \uparrow} \frac{d\omega}{2\pi \textrm{i}} \, [\,h^{(1)}_{\alpha\beta}+\omega\,\delta_{\alpha\beta}\, ]
\, g_{\beta\alpha}(\omega)
-  \frac{1}{2} \langle H^{(3)}\rangle \, ,
\label{eq_Koltun_W}
\end{equation}
where $h^{(1)}_{\alpha\beta}$ represents matrix elements (MEs) of the one-body part of the Hamiltonian and where the correlated three-body density matrix entering the last term is approximated via products of correlated one-body density matrices. 

Whereas the energy $E^A$ obtained from an unconstrained dSCGF($n$) calculation carries implicitly the intrinsic deformation $\beta^{\text{dSCGF}(n)}_2$  of the corresponding solution, the energy obtained from a constrained calculation based on the Routhian in Eq.~\eqref{eq_routhian} carries an explicit dependence on the constrained deformation $\beta^{\text{dSCGF}(n)}_2$ and is obtained via the modified  GMK sum rule
\begin{equation}
E^{A} = E'^{A} + \dfrac{1}{2}\,\mu_{q_{20}} \sum_{\alpha\beta} q^{(2 0)}_{\alpha\beta} ~\rho_{\beta\alpha} \, ,
\end{equation}
where $E'^{A}$ is obtained through Eq.~\eqref{eq_Koltun_W} using $H'$ instead of $H$.

\section{Numerical implementation}
\label{sec:numerics}

\subsection{Basis states and formal scaling}
\label{basisdep}

In actual calculations, all quantities of interest are represented starting from a one-body spherical harmonic oscillator (sHO) basis.
In this basis, all states up to a given $\emax \equiv \text{max}\{ 2n + \ell \}$ are included, where $n$ and $\ell$ denote the principal quantum number and the orbital angular momentum, respectively.
The total number of one-body basis states for a given $\emax$ is denoted as $\ndim \equiv \ndim(\emax)$.
While 1N and 2N operators are always represented in full, 3N operators are further limited to three-body basis states characterized by $\etmax \leq 3 \, \emax$.  Assessing and minimizing the error of physical observables with respect to both $e_\mathrm{1max}$ and $e_\mathrm{3max}$ constitutes a key aspect of any \ai{} nuclear structure calculation.

The ADC($n$) truncation scheme employed here leads to calculations that scale polynomially with $\ndim$, similarly to other state-of-the-art correlation-expansion methods~\cite{Hagen14,Hergert16,Tichai20b}.
The formal scaling for ADC($2$) and ADC($3$) is respectively $\ndim^5$ and $\ndim^6$~\cite{Banerjee23}.
In practice, this naive formal scaling is typically attenuated due to several considerations.
First, the various terms contributing to a given truncation, whose computing cost is generically said to scale as $\ndim^x$, actually scale as $A^y \, (\ndim-A)^{x-y}$\footnote{This refined scaling is due to the use of a Slater determinant reference state that partitions the one-body basis into the subsets of hole and particle states. When breaking U(1) symmetry via the use of a Bogoliubov state~\cite{Soma11,Signoracci15,Tichai18a}, such a refinement does not take place.}, with $\ndim \gg A$ and where $y$ depends on the considered term.
Second, when a given symmetry is enforced, the associated conservation law (parity, angular momentum, isospin) imposes selection rules such that many index combinations vanish. The working equations are thus sparse and can be recast in block-diagonal form, the dimension of each block being significantly smaller than in the symmetry-unrestricted version.

Beyond this reduction of the number of non-zero elements at fixed basis size, symmetry restrictions reduce the effective dimension of the working basis itself, with immense consequences on the cost of the calculation as soon as such symmetries must be relaxed to address open-shell nuclei.
Starting with a spherically-symmetric scheme for which SU(2) is enforced, one-body propagators at play are diagonal in the total angular momentum $j$ and its projection $m$, while being independent of the latter. This allows one to explicitly integrate over the $m$ quantum number ($J$-scheme implementation) and work in an effective basis characterized by a reduced set of quantum numbers, amounting to $\tndim =  \tndim(\emax)$ states, with $\tndim \ll \ndim$.
When rotational invariance is instead broken, one-body propagators do not carry a good $j$ anymore and are no longer degenerate in $m$. Thus, each value of $m$ has to be handled explicitly such that all sums at play must run over the full basis dimension $\ndim$ ($M$-scheme implementation) .
For all basis sizes, one has $\ndim \approx 10\,\tndim$~\cite{Frosini24}.
Given that SCGF calculations scale with powers of $\ndim$ as explained above, this translates into a computational effort that is orders of magnitude larger for dSCGF calculations than for spherical SCGF (sSCGF).
This is compounded by the significant growth of $\ndim$ with the mass number $A$ necessary to produce calculations characterized by a constant (in percentage) basis-size uncertainty.
These requirements constitute the main challenge of a deformed implementation and make the extension to heavy deformed nuclei a computational frontier for \ai{} many-body calculations.

\subsection{Approximate treatment of the 3N interaction}

Another important aspect of the numerical implementation relates to the treatment of 3N interactions. 
At the HF level, the 3N operator $H^{(3)}$ is included either exactly (such calculations are labeled ``Full 3N'') or through the rank-reduction (RR) procedure of Ref.~\cite{Frosini21}, which approximates $H^{(3)}$ into effective 0N, 1N and 2N operators.
While this approximation was shown to induce a 1-2\% error on total energies of light and medium-mass nuclei~\cite{Frosini21}, its performance in heavy deformed systems must eventually be assessed.
Given that (i) large $\etmax$ values become critical as one moves towards heavier systems and that (ii) dSCGF($2,3$) calculations must anyway rely on the RR procedure, establishing the validity of the RR in the heavy-mass region is critical. As a first step in this direction, such an assessment is presently performed in $^{28}$Si.

Since the contractions involving the 3N operator have to be re-evaluated at every dHF iteration, storing and decoupling 3N matrix elements in the full sHO basis quickly becomes computationally demanding, with both memory and CPU time acting as hard constraints. 
Consequently, dHF calculations in $^{28}$Si based on the full 3N interaction operator are restricted to $\etmax = 18$, which is the largest value that can be presently handled by the code employed here.
Deformed HF and SCGF calculations based on the RR approximation, by contrast, reach $\etmax = 24$\footnote{Values up to $\etmax = 28$ are accessible but unnecessary to converge calculations in a nucleus as light as $^{28}$Si.}.

Finally, let us mention that two other truncations affect matrix elements of 2N and 3N nuclear interactions prior to their transformation into the sHO basis employed in the many-body calculations~\cite{Hebeler21}.
The first truncation controls the number of total angular momentum components in the Jacobi 2N and 3N bases.
For \magic~\cite{Hebeler11a}, the main Hamiltonian used in the present study, and $\go$ used in Fig.~\ref{fig_SC}, these are limited respectively to $J^\text{2N}_\text{rel} \le 8$ and $J^\text{3N}_\text{rel} \le 15$. 
The second truncation, which only concerns the 2N part of the \magic interaction, is a basis cut $N_\text{max}$ imposed on the relative HO states used to perform the similarity renormalization group evolution of the Hamiltonian to lower resolution~\cite{Miyagi23}. In the present application $N_\text{max} \le 100$ has been used. While not done here, convergence studies  with respect to these truncation parameters will have to be performed in (heavy) deformed nuclei in the future.

\subsection{Numerical challenge}

The results presented in this work are obtained with \texttt{FoxTrot}, a newly developed many-body suite written for deformed \ai{} calculations, i.e.\ calculations where any spatial symmetry is explicitly allowed to break.
\texttt{FoxTrot} employs a spherical one-body, e.g.\ sHO, basis and implements dHF and dSCGF in $M$-scheme due to the fact that the one-body propagators at play are not assumed to be diagonal in the total angular momentum $J$, its projection $M$ or the parity $\Pi$ symmetry quantum numbers. The code reads 2N and 3N interaction MEs generated from \texttt{NuHamil}~\cite{Miyagi23} in the sHO basis and can perform the RR of the 3N interaction operator~\cite{Frosini21}. Furthermore, \texttt{FoxTrot} exploits a hybrid MPI/OpenMP parallelization to distribute memory and computational load simultaneously.

\begin{figure}
    \includegraphics[scale=0.70]{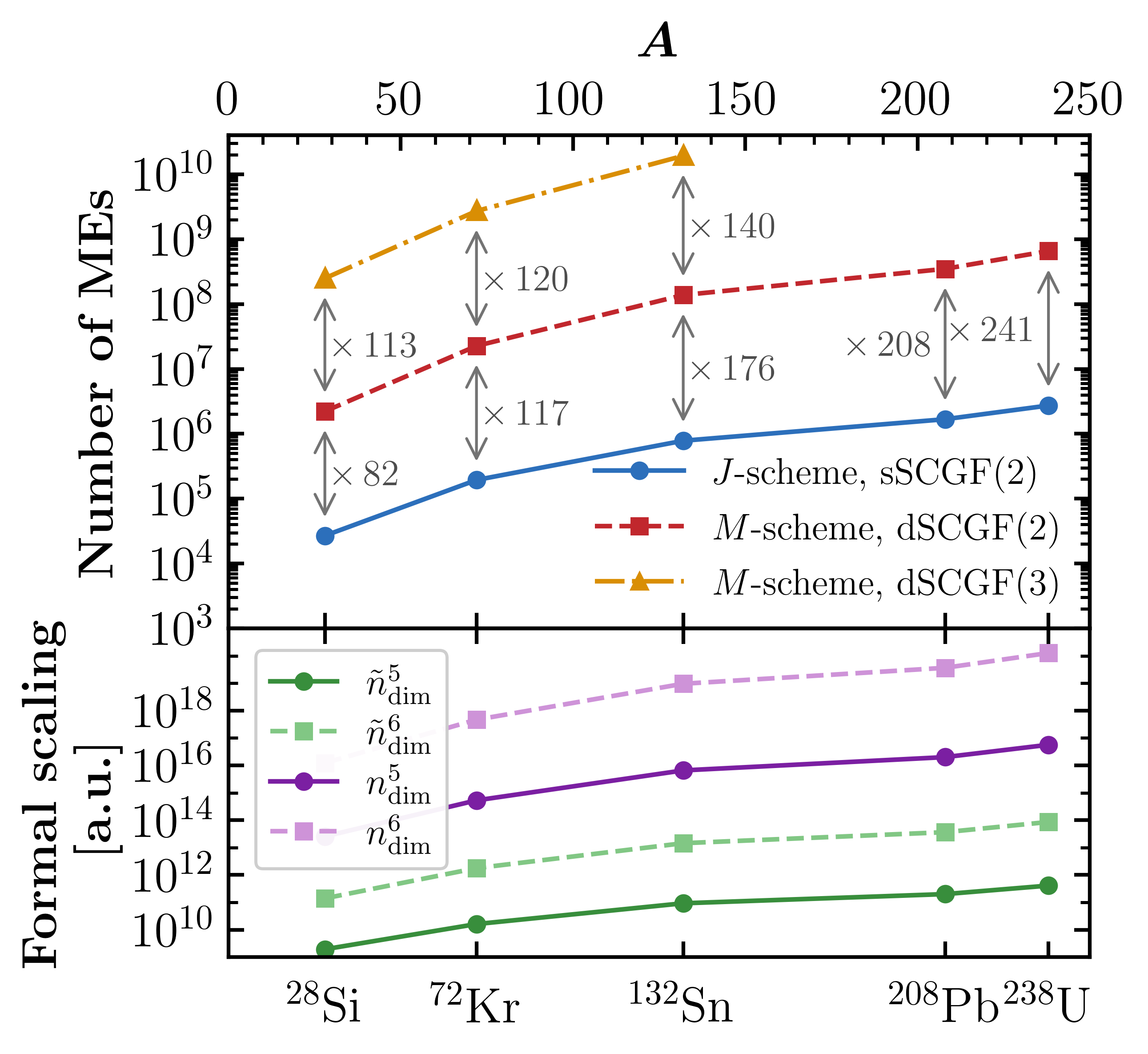}
    \caption{Comparison of the computational cost of $J$-scheme and axially-deformed $M$-scheme implementations of dSCGF($2,3$) calculations for five representative nuclei employing the \magic interaction.
    \emph{Upper panel}: number of non-zero matrix elements in the largest block of the ADC matrix.
    For each nucleus, $\emax$ is taken as the smallest value for which the dSCGF(2) total energy is converged to within $2\%$ respect to the extrapolated value.
    \emph{Lower panel}: formal CPU scaling evaluated for the number of one-body basis states corresponding to the $\emax$ used for each nucleus.}
    \label{NumMEs}
\end{figure}

Breaking rotational symmetry comes with a huge increase of the computational cost compared to spherically-symmetric calculations.
This is illustrated in Fig.~\ref{NumMEs} for five representative nuclei across the nuclear chart.
First, the upper panel compares the number of non-zero MEs entering the largest symmetry block of the ADC($2$) matrix in the two cases, i.e. in a $J$-scheme versus an axially-symmetric $M$-scheme implementation.
For a fixed basis-size error on the total energy of $2\%$\footnote{In the present example, the smallest basis dimensions needed to reach such a precision with the \magic interaction are $\emax = 7, 9, 11, 12$, and $13$ for $^{28}$Si, $^{72}$Kr, $^{132}$Sn, $^{208}$Pb and $^{238}$U respectively.}, the ratio between the two dimensions reaches about two orders of magnitude.
Going beyond second order further complicates the picture: while the $E^>+C$ and $E^<+D$ blocks of the ADC($2$) matrix are diagonal, at the ADC($3$) level they are full (although sparse).
This leads to an increase of two additional orders of magnitude in the number of stored MEs, as visible in Fig.~\ref{NumMEs}.
While these MEs can be distributed across MPI ranks in order to contain the memory footprint per node, the large total memory cost constitutes a significant challenge for the numerical implementation.

The lower panel of Fig.~\ref{NumMEs} compares the formal scaling in $J$-scheme versus $M$-scheme  associated with the computation of the costliest MEs of the ADC($2,3$) matrix. Because the ratio $n_{\mathrm{dim}}/\tilde{n}_{\mathrm{dim}}$ is about one order of magnitude for all values of $\emax$ (see Tab.~\ref{tab:dim}), going from sSCGF($2$) (sSCGF($3$)) to dSCGF($2$) (dSCGF($3$)) amounts to an increase of about five (six) orders of magnitude in CPU time.

Combining the information from both panels, the nominal increase in CPU time at second order when going from $J$-scheme to $M$-scheme is two orders of magnitude from the number of MEs and five from their actual computation, for a total of seven orders of magnitude.
Moving to dSCGF($3$) calculations generates another two orders of magnitude more MEs and makes their actual computation one order of magnitude costlier, for a total of an extra three orders of magnitude compared to  dSCGF($2$) calculations. Once the ADC matrix is built, it must be diagonalized. While this step carries a non-negligible computational cost, it is largely reduced by the Krylov projection, such that the overall cost remains dominated by the evaluation of the matrix elements entering the ADC matrix.

\begin{table}
  \begin{ruledtabular}
    \begin{tabular}{cccccccc}
      $\emax$                    &   7 &   8 &   9 &   10 &   11 &   12 &   13 \\
      \colrule
      \noalign{\vspace{3pt}}
      $\tilde{n}_{\mathrm{dim}}$ &  72 &  90 & 110 &  132 &  156 &  182 &  210 \\
      $n_{\mathrm{dim}}$         & 480 & 660 & 880 & 1144 & 1456 & 1820 & 2240 \\
    \end{tabular}
      \caption{\label{tab:dim}%
    Number of one-body basis states to be explicitly handled in $J$-scheme ($\tilde{n}_{\mathrm{dim}}$) and in $M$-scheme ($n_{\mathrm{dim}}$) as a
    function of $\emax$.}
  \end{ruledtabular}
\end{table}

Going from light to heavy nuclei comes in itself with a pronounced increase of the computational cost. This is also quantified in Fig.~\ref{NumMEs}. For a fixed basis-size error ($2\%$) on the total energy, the number of stored MEs in the $M$-scheme ADC($2$) matrix increases by slightly more than one (two) order(s) of magnitude when going from $^{28}$Si to $^{132}$Sn ($^{238}$U). At the ADC($3$) level, an increase by two orders of magnitude is visible  going from $^{28}$Si to $^{132}$Sn. As for the computation cost of those MEs, the necessity to augment $\emax$ to maintain the same basis-size error on the total energy leads to an increase of about three (four) orders of magnitude in CPU time at the ADC($2$) (ADC($3$)) level when going from $^{28}$Si to $^{238}$U. Combining the information from both panels, this makes dSCGF($2$) (dSCGF($3$)) calculations of very heavy deformed nuclei about five (seven) orders of magnitude more expensive than in light deformed nuclei. By comparison, the above naive analysis indicates that the computation of the doubly open-shell $^{238}$U nucleus is about eight (eleven) orders of magnitude more expensive at the ADC($2$) (ADC(3)) level than the computation of the doubly closed-shell $^{208}$Pb nucleus.

\subsection{CPU time and memory costs}

Several complementary numerical strategies allow the dSCGF implementation in \texttt{FoxTrot} to reach heavy masses without facing a memory usage or CPU time bottleneck.
First, the effective 2N interaction is never stored in $M$-scheme; it is instead kept in the compact $J$-scheme representation and decoupled on the fly.
Second, diagrammatic contributions to the self-energy~\cite{Raimondi18} are evaluated in an order that minimizes the number of times a given decoupling must be performed.
Third, the ADC matrix is built one symmetry block at a time.
For each block, the two sub-matrices $E^> + C$ and $E^< + D$ are constructed separately and their dimension is immediately reduced by a Krylov projection.
The $M$ and $N$ matrices are subsequently Krylov-projected as well.
In this way, one never needs to store multiple copies of the full blocks of the Dyson matrix, but only their memory-inexpensive Krylov projections.

\begin{figure}
    \includegraphics[scale=0.68]{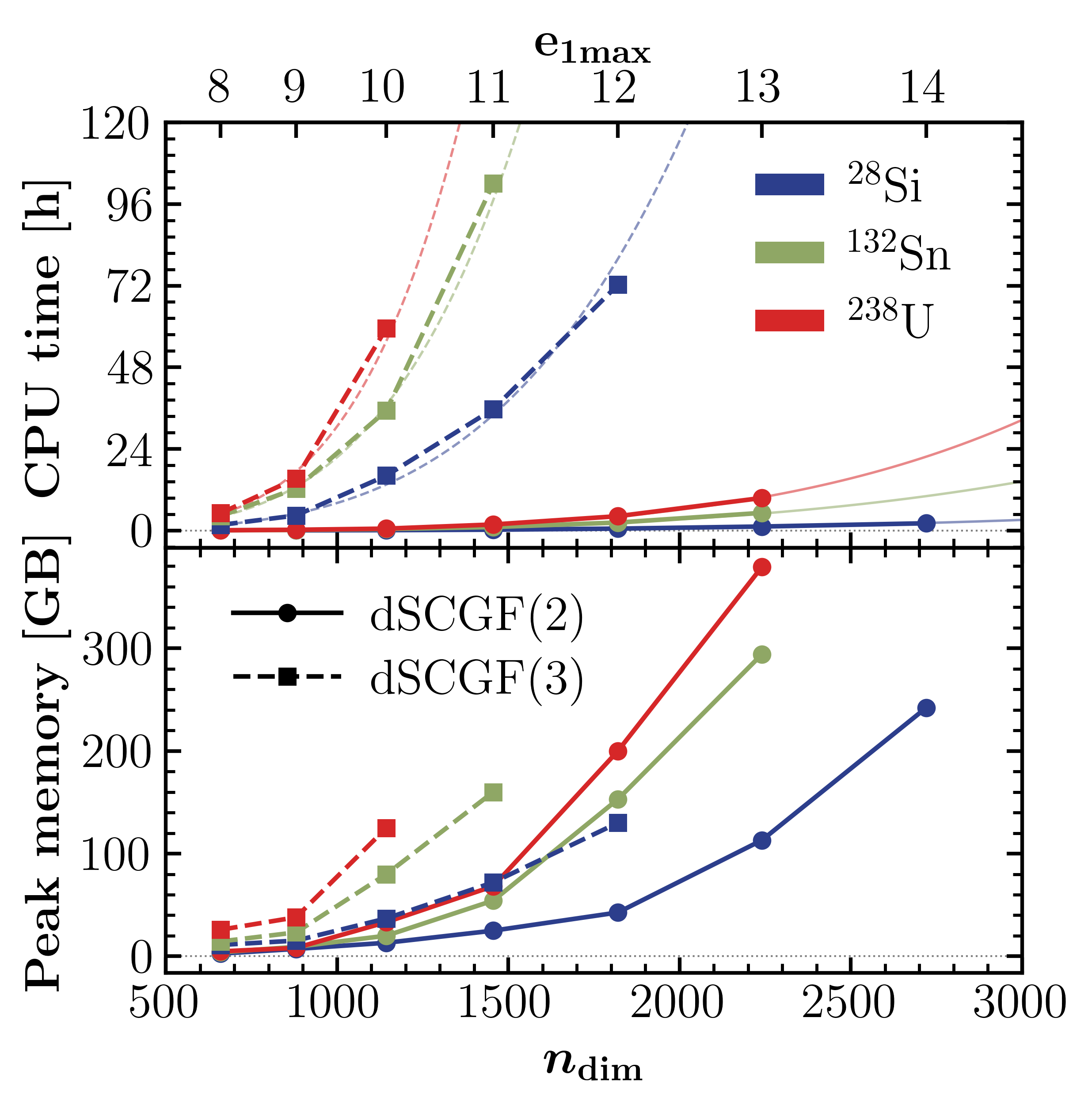}
    \caption{Cost of dSCGF($2,3$) calculations of $^{28}$Si, $^{132}$Sn and $^{238}$U for increasing values of $\emax$ ($n_{\mathrm{dim}}$). \emph{Upper panel}: CPU time in hours. \emph{Lower panel}: Peak RAM memory in GB. All calculations have been performed with 1 MPI process and 32 OpenMP threads, on nodes equipped with two 32-core AMD EPYC 9354 processors (Zen 4, 3.25 GHz).}
    \label{TimeMemory}
\end{figure}

Figure~\ref{TimeMemory} collects the actual CPU time and memory required for dSCGF($2,3$) calculations of $^{28}$Si, $^{132}$Sn and $^{238}$U as a function of $\emax$ ($n_{\mathrm{dim}}$).
From the upper panel, one observes that the increase is very mild for dSCGF($2$), while it becomes much steeper for dSCGF(3).
While the figure shows results obtained with a single MPI process, it is worth noting that the multi-node parallelization of \texttt{FoxTrot} scales very efficiently with the number of processes, such that using four MPI processes reduces the computational time by almost a factor of four.
The many-body solver is therefore not the limiting factor to push calculations to larger bases: the code can run calculations beyond $\emax=14$ such that the present limitations are dictated by the generation and storage of 3N MEs at large $\etmax$.

The data appearing in the upper panel of Fig.~\ref{TimeMemory} have been fitted and extrapolated according to the scaling law
\begin{equation}
y = C A^q n_{\rm dim}^p ,
\label{eqn:scalingAp}
\end{equation}
where the explicit A dependence refines the naive scaling according to the discussion provided in Sec.~\ref{basisdep}. Results are gathered in Tab.~\ref{tab:scaling-fit}.
\begin{table}[b]
\centering
\begin{ruledtabular}
\begin{tabular}{lccc}
Method & $C$ & $q$ & $p$ \\
\colrule
\noalign{\vspace{3pt}}
dSCGF(2) & $1.66 \times 10^{-14}$ & $0.89 \pm 0.04$ & $3.76 \pm 0.08$ \\
dSCGF(3) & $1.26 \times 10^{-12}$ & $0.61 \pm 0.04$ & $3.97 \pm 0.12$ \\
\end{tabular}
\caption{Fitted parameters of the scaling law of Eq.~\eqref{eqn:scalingAp} for dSCGF($2,3$) calculations based on data appearing in Fig.~\ref{TimeMemory}.}
\label{tab:scaling-fit}
\end{ruledtabular}
\end{table}
The exponents $q$ and $p$ are essentially unchanged when going from dSCGF($2$) to dSCGF($3$) calculations and are such that $p+q\approx 4.6$. This indicates a very similar effective scaling with respect to both the mass of the system and the basis size, in a way that makes actual dSCGF($2$) (dSCGF($3$)) calculations (much) more gentle than what was expected from the naive scaling discussed earlier in conjunction with the lower panel of Fig.~\ref{NumMEs}.
Eventually, the increased computational cost of dSCGF($3$) calculations is almost entirely absorbed in the prefactor $C$, which grows by nearly two orders of magnitude with respect to dSCGF($2$).

The lower panel of Fig.~\ref{TimeMemory} displays the peak memory usage for the calculations shown in the upper panel.
Going from dSCGF($2$) to dSCGF($3$) calculations, a sharp increase is visible in connection with  the non-zero off-diagonal MEs appearing in the $C$ and $D$ blocks as discussed earlier.
The MPI parallelization helps control the growth of the number of MEs with $\emax$ by distributing them over different nodes.
The increase of memory with $A$ is also evident: at fixed $\emax=13$, calculations require about 100~GB for $^{28}$Si, less than 300~GB for $^{132}$Sn, and nearly 400~GB for $^{238}$U.

The favorable scaling and moderate memory footprint of actual calculations based on the implementation in \texttt{FoxTrot} discussed above bring heavy deformed systems within numerical reach, which will constitute the focus of future applications.

\section{Application to $^{28}$Si}
\label{28SiAnalysis}

In order to illustrate the rationale of dSCGF calculations and associated results, the case of the light $^{28}$Si nucleus is considered in the present work.  This system has been studied extensively both experimentally~\cite{Endt:1973div} and theoretically~\cite{Frycz24,Taniguchi:2026hnj}. In particular, the ground state has been associated with an oblate intrinsic structure whereas a $J^\pi=0^+$ excited state at $E_\text{x} = 6.69$\,MeV has been characterized as a probable prolate shape isomer~\cite{kanada03}.

Calculations are presently performed using the \magic Hamiltonian~\cite{Hebeler11a}.

\subsection{Impact of the 3N force on the dHF TEC}

A key aspect of \ai{} calculations of doubly open-shell nuclei based on expansion methods employing a dHF reference state relates to the behavior of the energy of the latter as a function of its axial quadrupole deformation $\beta^{\text{dHF}}_2$, i.e.\ to the behavior of the dHF TEC obtained via constrained dHF calculations. In particular, one is presently interested in characterizing the behavior of the TEC both as a function of $\emax$ and as a function of $\etmax$ for fixed $\emax$. In addition, one wishes to quantify the effect of the RR operated on the 3N interaction that is currently mandatory to perform correlated calculations beyond the mean field. 

\begin{figure}
	\includegraphics[scale=0.70]{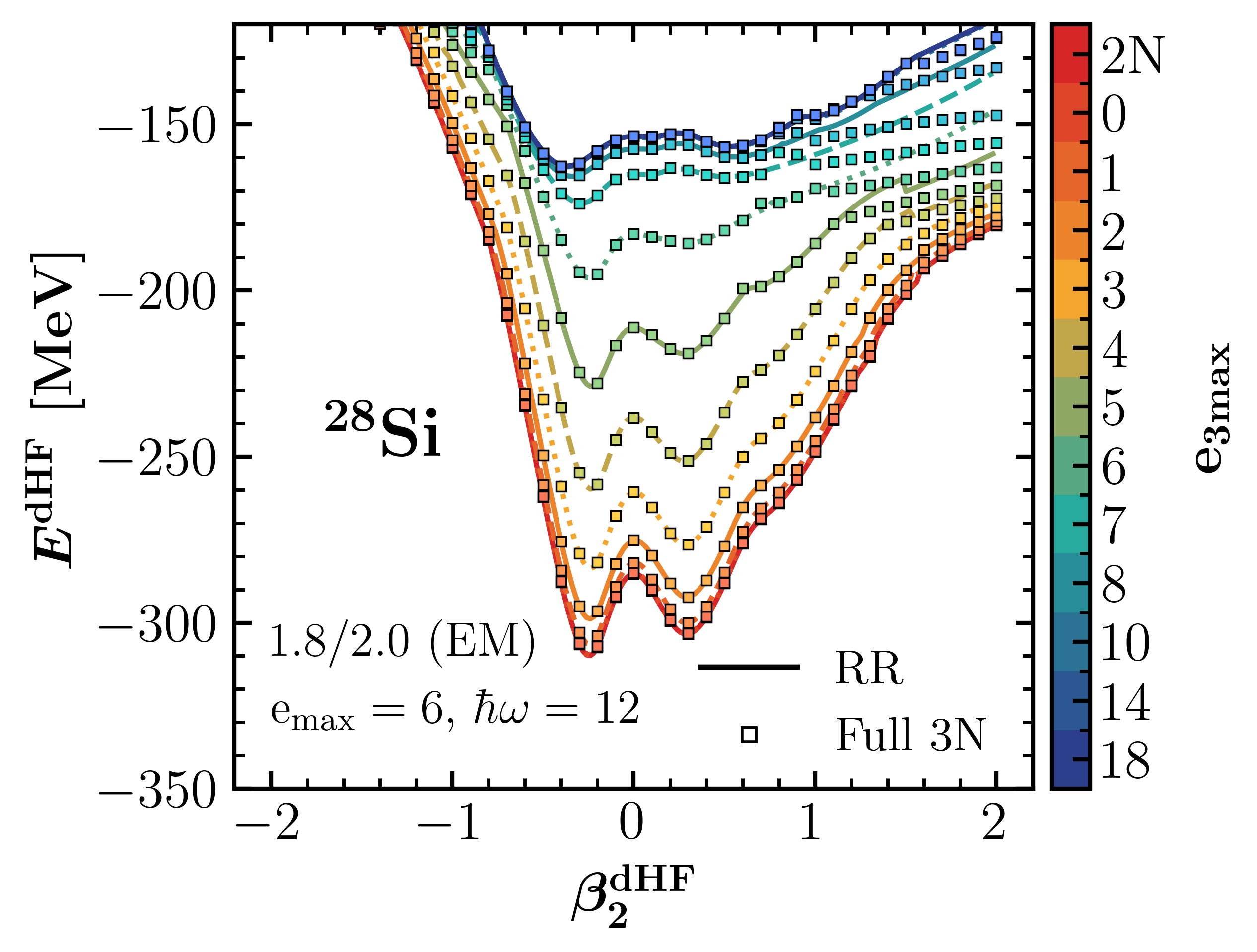}
	\caption{Deformed Hartree-Fock TEC of $^{28}$Si as a function of the intrinsic dHF quadrupole deformation $\beta^{\text{dHF}}_2$ computed for different $\etmax$ truncations. Results obtained with the full 3N operator and its RR version are compared. Calculations based on the \magic Hamiltonian are performed with $\emax=6$ and $\hbar \omega = 12$\,MeV.}
	\label{fig_reduction_Si}
\end{figure}

Figure~\ref{fig_reduction_Si} displays the dHF TEC of $^{28}$Si obtained using $(\hbar\omega,\emax)=(12\,\text{MeV},6)$ and the full 3N interaction operator or its RR approximation\footnote{The employed sHO frequency $\hbar\omega=12$\,MeV is optimal at the dHF level and the TEC is largely converged at $\emax=6$ on the scale employed in Fig.~\ref{fig_reduction_Si}.}. Results are shown for the 2N interaction alone, as well as for $\etmax$ ranging from $0$ all the way to $3\, \emax=18$, i.e.\ using the uncut set of 3N
MEs.

One first observes that the dHF TECs computed with the full 3N interaction operator and its RR version are essentially identical except at very large intrinsic deformations ($\beta^{\text{dHF}}_2 \ge 1$) for intermediate $\etmax=6-10$ values. In such a regime, the TEC obtained from the genuine 3N interaction operator flattens out at large deformation whereas the RR delivers increasingly stiffer TECs. Further increasing $\etmax$, the TEC computed from the full 3N interaction operator stiffens again and eventually catches up with the maximally stiff TEC delivered by the RR for $\etmax = 3\,\emax$. This indicates that the RR itself performs well across the entire TEC, as long as $\etmax$ is close enough to its maximum value. These results support the use of the RR procedure in the calculation of deformed nuclei as long as  $\etmax$ values close to $3\, \emax$ can be handled in the first place to perform the RR. The latter requirement will of course become challenging to fulfill in heavy nuclei where significantly larger $\emax$ and thus $\etmax$ values are necessary to achieve converged enough calculations.

While the TEC obtained with the sole 2N interaction is qualitatively at odds with the one obtained from phenomenological, i.e.\ EDF, calculations~\cite{AMEDEEDATABASE}, the gradual inclusion of the 3N interaction (i.e., going from red to blue in Fig.~\ref{fig_reduction_Si}) makes the TEC consistent with the latter. The 3N interaction achieves this by providing a large repulsive effect that tends to strongly decrease with intrinsic deformation, even though that trend displays a non-trivial evolution with $\etmax$, especially at large prolate deformations as discussed above. This eventually results in a TEC that becomes much softer with deformation and varies over a phenomenologically meaningful scale: the topology around the spherical point is such that the intrinsic deformations of the oblate and prolate minima increase to  $\beta^{\text{dHF}}_2 \approx -0.35$ and  $\beta^{\text{dHF}}_2 \approx 0.5$, respectively, while the barrier between them decreases from $\sim 30$ MeV down to a few MeV.

\subsection{dSCGF(2,3) solutions}

\begin{figure}
	\includegraphics[scale=0.70]{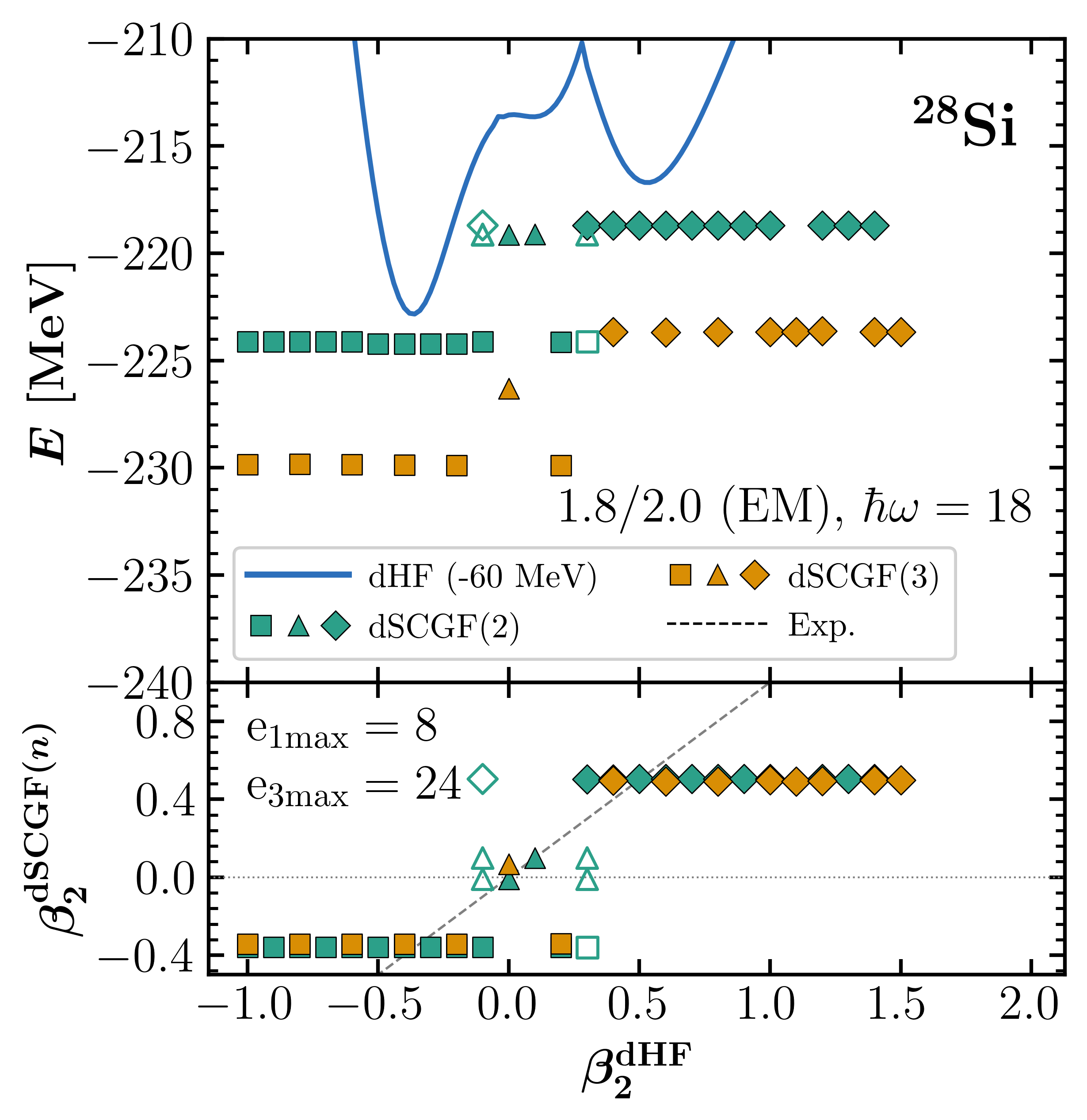}
	\caption{Results corresponding to four dSCGF(2,3) solutions obtained in $^{28}$Si while scanning the intrinsic deformation $\beta^{\text{dHF}}_2$ of the mean-field dHF starting state. Calculations employ the \magic Hamiltonian and $(\hbar\omega, \emax, \etmax)= (18\,\text{MeV},8,24)$. Different symbols are used for the four different solutions. Full symbols designate the solutions found via an unconstrained single run algorithm. Open symbols specify additional solutions found from a single starting dHF state via a refined algorithm including a constraint on the {\it correlated} intrinsic deformation $\beta^{\text{dSCGF}(n)}_2$ for a few SCGF iterations before relaxing it. \textit{Upper panel}: energies against the (shifted) dHF TEC.
    \textit{Lower panel}: intrinsic deformation $\beta^{\text{dSCGF(2,3)}}_2$ of the four solutions.}
	\label{fig_ADC_Si}
\end{figure}

The dHF TEC discussed above is a crucial ingredient because dHF propagators constitute the starting point of dSCGF(2,3) calculations. As explained in Sec.~\ref{sec:dscgf}, selecting a (global or local) minimum in the dHF TEC constitutes a natural option to initiate the dSCGF calculation. Still, any point on the dHF TEC can in fact be chosen as a starting point. Indeed, and as also explained in Sec.~\ref{sec:dscgf}, the solution of {\it self-consistent} Green's function calculations holds the promise to be {\it independent} of the starting dHF state at any ADC($n$) truncation order\footnote{The unconstrained dSCGF(1)$\equiv$dHF solution corresponding to the absolute minimum of the dHF TEC is itself independent of the starting point of the calculation.}. As an illustrative example, Fig.~\ref{fig_ADC_Si} displays the dSCGF($2$) and dSCGF($3$) energies obtained for $(\hbar\omega, \emax, \etmax)= (18\,\text{MeV},8,24)$ starting from all dHF states along the TEC, knowing that $\hbar\omega=18$\,MeV was checked to constitute the optimal sHO frequency at the dSCGF($2$) level. 

 Despite the fact that the starting states along the dHF TEC differ by as much as 70 MeV, dSCGF($2$) energies converge to only four distinguishable values differing at most by a few MeV. Those solutions are located about 60\,MeV below the dHF minima thanks to the energy gain brought by the inclusion of dynamical correlations beyond the deformed mean field. Two of the solutions are nearly degenerate and carry very small intrinsic quadrupole deformations $\beta^{\text{dSCGF}(2)}_2$ (see bottom panel). These two solutions do not however describe physical states given that they disappear at larger $\emax$ values. Contrarily, the oblate solution at $\beta^{\text{dSCGF(2)}}_2 \approx -0.35$ is identified with the $^{28}$Si ground state whereas the prolate solution at $\beta^{\text{dSCGF(2)}}_2 \approx 0.5$ is identified with a shape isomer. It is clear that, independently of the starting dHF energy and deformation, the self-consistent iterations drive the dSCGF($2$) solution to a small possible set of physical solutions carrying specific energies and intrinsic deformations\footnote{While not shown here, deformed many-body perturbation theory (dMBPT) calculations~\cite{Frosini21} keep a strong memory of the initial dHF state intrinsic deformation due to the non self-consistent character of the method. As a result, straight dMBPT calculations performed on top of the dHF TEC generate a non-trivial TEC.}. Eventually moving to the more advanced dSCGF($3$) level, the oblate and prolate solutions retain essentially the same intrinsic deformation while gaining $5.7$ and $4.9$\,MeV, respectively. 

As explained in Sec.~\ref{sec:dscgf}, dSCGF($n$) solutions based on a dHF state breaking rotational symmetry would exactly restore that symmetry in the exact limit ($n\rightarrow \infty$). In practice, restoring a broken symmetry requires summing specific infinite-order contributions when expanding an eigenstate in terms of particle-hole configurations on top of a symmetry-breaking reference state as in dSCGF theory, dMBPT or deformed coupled cluster (dCC) theory~\cite{Novario20}. Consequently, low-order truncation schemes cannot effectively restore a broken symmetry and doing so requires the inclusion of an explicit symmetry projection technique~\cite{Duguet15a}. While this has been formulated for dMBPT or dCC theory~\cite{Duguet15a,Qiu17,Duguet17a,Qiu18}, and implemented for the latter~\cite{Hagen22}, it remains to be done for dSCGF theory. In the absence of an explicit angular-momentum projection, a dSCGF($n$) solution with $\beta^{\text{dSCGF}(n)}_2\neq 0$ describes an {\it intrinsic} state associated with a sequence of $J^{\pi}=0^+, 2^+, 4^+\ldots$, i.e.\  a linear combination of such a set of eigenstates. While the angular-momentum projection is key to resolve each member of the rotational band, the intrinsic state already delivers an excellent approximation of bulk properties of the $J^{\pi}=0^+$ bandhead~\cite{Frosini22b,Frosini22c,Hagen22,Hu:2024jih,Sun:2024iht,Bofos:2026nmg}, e.g.\ the correction to the ground-state energy brought by the symmetry restoration is largely subleading, especially in heavy systems.

As seen in Fig.~\ref{fig_ADC_Si}, performing an unconstrained calculation at each $\beta^{\text{dHF}}_2$ value, the dSCGF calculation converges towards one of the identified solutions (full symbols). Most often, unconstrained runs based on dHF states with similar deformations tend to reach the same dSCGF solution, implying an apparent link between a specific subset of dHF states and a given final solution. This seems to contradict the statement that dSCGF solutions are independent of the starting point. While this apparent link is fictitious, it reflects that a given dHF state entertains a larger adiabatic connection to a given correlated solution that will thus be preferably reached. Still, every dHF starting point can in fact lead to any final eigenstate that is not orthogonal to it. By refining the numerical algorithm, the four final solutions appearing in Fig.~\ref{fig_ADC_Si}  can indeed be reached starting from {\it any} of the dHF states along the TEC. This is illustrated in Fig.~\ref{fig_ADC_Si} for the two (arbitrary) dHF states at $\beta^{\text{dHF}}_2 \approx -0.1$ and $0.3$. Beyond the solution found through the straight unconstrained run (full symbol), the other three solutions (open symbols) are indeed accessed in each case by imposing a constraint on the correlated intrinsic deformation $\beta^{\text{dSCGF}(2)}_2$ for a few SCGF iterations before relaxing it to let the self-consistent solution converge freely to a given correlated solution. By scanning an appropriate range of $\beta^{\text{dSCGF}(2)}_2$ values, one can indeed manage to reach all four solutions starting from any dHF state along the TEC.

The above findings are a clear illustration of (i) the \textit{self-consistent} character of dSCGF calculations that make correlated propagators eventually independent of the dHF starting point and (ii) the capacity to reach different solutions of the Schr\"odinger equation at once by scanning either the intrinsic deformation of the initial dHF state or the one of the final correlated state.

\begin{figure}
	\includegraphics[scale=0.70]{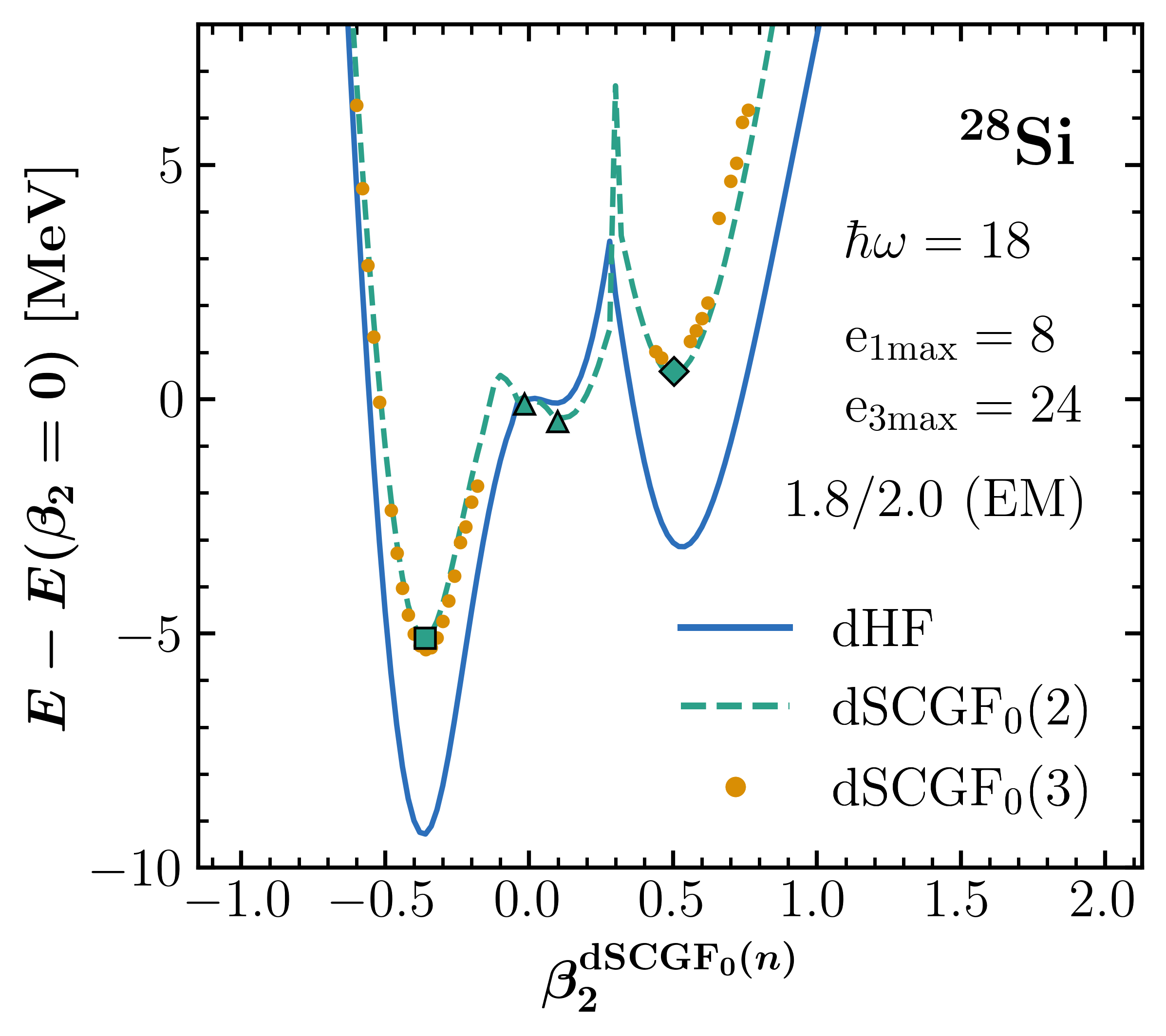}
	\caption{Total energy curve in $^{28}$Si at the dHF and  dSCGF$_0$(2,3) levels as a function of the corresponding intrinsic deformation $\beta^{\text{dSCGF}(n)}_2$. The three curves are shifted vertically by setting their value at $\beta^{\text{dSCGF}(n)}_2=0$ to zero. Calculations employ the \magic Hamiltonian and $(\hbar\omega, \emax, \etmax)= (18\,\text{MeV},8,24)$. The four solutions obtained through unconstrained dSCGF$_0$(2) calculations are further indicated as full green symbols. }
	\label{fig_ADC_SiTEC}
\end{figure}

Independently of the starting dHF state, the refined algorithm described above does underline the utility of scanning the intrinsic deformation of the correlated solutions to efficiently identify them, thus giving credit to the notion of {\it correlated} TEC obtained via constrained dSCGF($n$) calculations. In Fig.~\ref{fig_ADC_SiTEC}, the unconstrained dSCGF$_0(2)$ solutions\footnote{For simplicity, results displayed in Fig.~\ref{fig_ADC_SiTEC} originate from dSCGF$_0(2)$ calculations rather than from fully self-consistent dSCGF($2$) calculations. It has been checked for many points along the TEC that dSCGF($2$) solutions are only between $0.1$ and $0.5$\,MeV below dSCGF$_0(2)$ ones.} identified earlier are superimposed onto the dSCGF$_0(2)$ TEC displayed as a function of the correlated intrinsic deformation. The dHF and dSCGF$_0(3)$ TECs are also displayed as a function of the corresponding intrinsic deformations, knowing that all three TECs are shifted vertically by setting their value at $\beta^{\text{dSCGF}(n)}_2=0$ to zero. One first observes that the four unconstrained dSCGF$_0(2)$ solutions appear as minima of the constrained dSCGF$_0(2)$ TEC, giving credit to the variational character of dSCGF($2$) calculations associated with an underlying $\Phi$-functional~\cite{Soma11}. The key additional observations that can be made are
\begin{enumerate}
\item The dSCGF$_0(2)$ TEC displays one more minimum than the dHF one, illustrating the fact that it is possible to have a different number of solutions at various dSCGF($n$) orders. Eventually, it is important to verify the stability of the obtained solutions with respect to both $(\emax,\etmax)$ values and the many-body truncation order.
\item Regarding the two viable minima for which a one-to-one connection can be safely performed, dSCGF$_0(2,3)$ solutions keep a strong memory of dHF ones, i.e.\ the intrinsic deformations of the former are very close to (i.e.\ only slightly smaller than) the intrinsic deformations of the latter.
\item Even though the three TECs look similar, dynamical correlations do modify the topology of the TEC: the dSCGF$_0(2)$ TEC is overall stiffer with respect to intrinsic deformation than the dHF one, making both deformed minima move up relative to the near-spherical one(s). As a matter of fact, the positions of the prolate and near-spherical minima are inverted at the dSCGF$_0(2)$ level compared to the dHF results. The same remains true for the dSCGF$_0(3)$ TEC\footnote{The portion of the dSCGF$_0(3)$ TEC around $\beta^{\text{dSCGF}(3)}_2=0$ is missing due to the difficulty to converge the dSCGF$_0(3)$ loop in this region. This is due to the fact that $M$ and $N$ matrices include coupled-cluster amplitudes whose denominators can be very close to zero. Producing a fully self-consistent dSCGF($3$) TEC is required to overcome this limitation but was not performed here for simplicity.} whose topology is close to the  dSCGF$_0(2)$ one.
\end{enumerate}

\subsection{Binding energy and excitation spectrum}

\begin{figure}
	\includegraphics[scale=0.60]{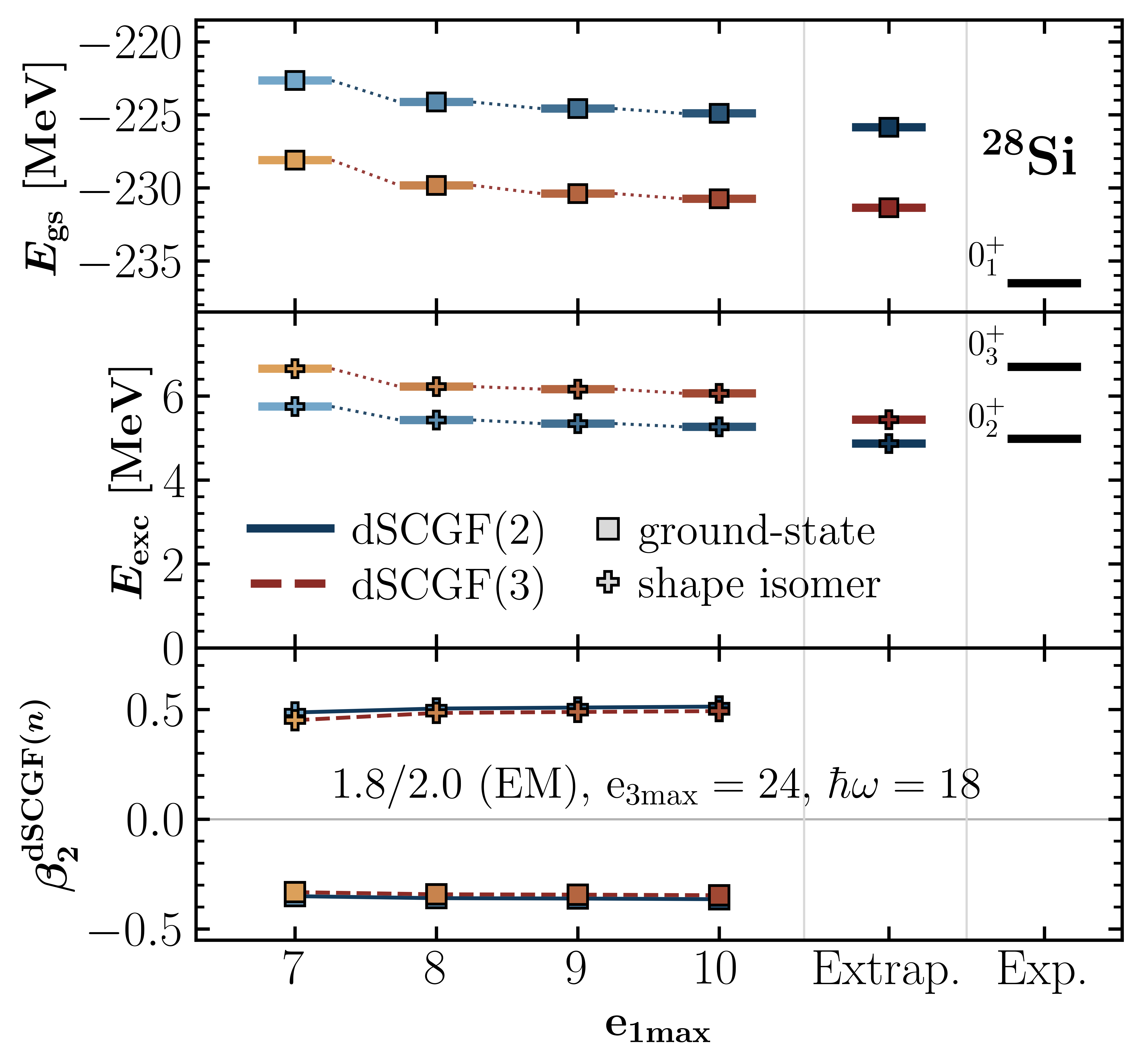}
	\caption{Results of dSCGF(2,3) calculations in $^{28}$Si as a function of $\emax$ based on the \magic Hamiltonian. Calculations employ $\hbar\omega= 18$\,MeV and $\etmax= 24$. \textit{Upper panel}: ground-state energy  from $\emax = 6$ to $\emax = 10$  plus the extrapolated value to $\emax = \infty$ along with the experimental binding energy. \textit{Middle panel}: excitation energy of the prolate shape isomer from $\emax = 6$ to $\emax = 10$ against experimentally known $J^{\pi}=0^+$ excited states below 8\,MeV. \textit{Bottom panel}: intrinsic axial quadrupole deformation of the two dSCGF(2,3) solutions.}
	\label{fig_convergence_Si}
\end{figure}

The picture depicted in Fig.~\ref{fig_ADC_Si} for the oblate and prolate solutions is found consistently at successive $\emax$ values. Hence, the energy convergence of the solutions can be tracked and investigated. Figure~\ref{fig_convergence_Si} displays the convergence with $\emax$ (for fixed $\etmax=24$) of the ground-state energy associated with the oblate solution and of the excitation energy associated with the  prolate solution, for both dSCGF($2$) and dSCGF($3$) calculations. Energies extrapolated to the infinite basis-size limit ($\emax\rightarrow\infty$) are compared to those of the three lowest $J^{\pi}=0^+$ states known experimentally. Energies are extrapolated by fitting their $\emax$ dependence with the three-parameter exponential ansatz $E(\emax) = E_\infty + a\,e^{-k\,\emax}$~\cite{Bogner:2007rx}, where $E_\infty$, $a$ and $k$ are determined by a least-squares fit.

One first observes from the upper panel of Fig.~\ref{fig_convergence_Si} that the binding energy  shows a converging pattern with $\emax$ such that essentially converged values are in fact already obtained at $\emax=10$. The bottom panel shows that the corresponding ground-state intrinsic deformation is very stable with $\emax$ and does not change going from dSCGF($2$) to dSCGF($3$) calculations. While the extrapolated dSCGF($2$) binding energy underestimates experiment by $4.9\%$, the extra few MeVs gained at the dSCGF($3$) level reduce the discrepancy to about $2.1\%$. This is consistent with results obtained in this mass region with the \magic Hamiltonian~\cite{Stroberg21}.

The middle panel shows that dynamical correlations added at the dSCGF($3$) level raise the excitation energy of the intrinsically prolate solution from $4.87$\,MeV to $5.44$\,MeV. The intrinsic prolate deformation of the state is very stable with $\emax$ and identical at the dSCGF($2$) and dSCGF($3$) levels. While strong experimental evidence is missing, many phenomenological calculations consistently indicate that the $0^+_3$ state is a convincing candidate for a prolate shape isomer, e.g.\ see Refs.~\cite{kanada03,Frycz24,Taniguchi:2026hnj} and references cited therein. The dSCGF($3$) excitation energy of the prolate solution, extrapolated to the infinite basis-size limit, presently obtained based on the \magic Hamiltonian is in fair agreement with experimental data, i.e.\ it underestimates the excitation energy of the experimental $0^+_3$ state ($E_\text{x} = 6.69$\,MeV) by about $1.2$\,MeV.

\section{Conclusions}
\label{sec:conclusions}

The self-consistent Green's function formalism has been extended to doubly open-shell nuclei by allowing the one-body propagator to spontaneously break SU(2) rotational symmetry.
The resulting deformed SCGF scheme is based on the ADC expansion of the self energy and has been implemented in the newly developed \texttt{FoxTrot} many-body suite at first, second, and third order truncation levels. While the implementation is symmetry-unrestricted by design, the present applications have been limited to parity-conserving axial deformations.

The numerical price to pay for breaking rotational symmetry has been quantified in detail.
Working in $M$-scheme at dSCGF($2$) (dSCGF($3$)) level increases the number of MEs to be computed by two (four) orders of magnitude with respect to a second-order (third-order) $J$-scheme implementation. Given the naive $\ndim^5$ ($\ndim^6$) scaling of dSCGF($2$) (dSCGF($3$)) calculations, this translates into seven (ten) orders of magnitude in computational cost. This tremendous increase is leveraged thanks to several numerical strategies. First, the effective two-nucleon interaction stored in $J$-scheme representation is decoupled on the fly, knowing that self-energy diagrams are evaluated in an order that minimizes the number of such decouplings. Second, dimensionality-reduction techniques are employed to strongly reduce both the memory footprint and CPU time. Namely, (i) on-the-fly Krylov projection of individual symmetry blocks of the $E^> + C$, $E^< + D$, $M$ and $N$ sub-matrices of the ADC matrix is implemented and (ii) an optimized reference state is exploited to compress the information contained in the dSCGF($2,3$) propagator into an effective mean-field-like one at the start of each self-consistent iteration. Third, the $C$ and $D$ blocks of the ADC matrix are distributed across MPI ranks. All this contributes to keeping both the memory footprint and the CPU time within reach even for large basis dimensions amenable to the computation of (very) heavy nuclei. In fact, the many-body solver is not the limiting factor in this respect: the present frontier is set by the generation and storage of three-nucleon interaction matrix elements necessary to perform converged calculations of (very) heavy systems.
 
The rationale and outcome of dSCGF($2,3$) calculations have been illustrated through a detailed study of $^{28}$Si based on the \magic Hamiltonian, from which several general lessons emerge.
 
First, the rank-reduction approximation applied to the three-nucleon interaction operator, which is mandatory beyond the mean-field level, has been benchmarked against the explicit treatment along the entire dHF total energy curve.
The two agree closely except at large prolate deformations for intermediate $\etmax$ values, the discrepancy vanishing as $\etmax$ approaches $3\,\emax$.
This validates the use of the rank reduction, provided $\etmax$ values close to $3\emax$ can be handled in the first place.
 
Second, the genuinely self-consistent character of the method has been demonstrated explicitly.
Scanning the intrinsic deformation of the dHF starting state, dSCGF($2$) calculations converge to a small number of distinct solutions, and each of them can be reached from \emph{any} point along the dHF total energy curve once a transient constraint on the correlated deformation is imposed and subsequently relaxed.

Third, constrained dSCGF($n$) calculations deliver a \emph{correlated} total energy curve in which the unconstrained solutions appear as minima.
Dynamical correlations beyond the deformed mean-field do modify the topology of the TEC: the correlated curve is overall stiffer than the dHF one such that minima carrying large intrinsic deformations are pushed up in energy relative to less deformed ones at the dSCGF level. 
Consequently, the existence and energy of correlated solutions cannot be easily inferred from the mean-field topology alone.
 
Fourth, tracking the two physical solutions with $\emax$ in $^{28}$Si shows that the extrapolated dSCGF($2$) binding energy underestimates experiment by $4.9\%$, a discrepancy reduced to $2.1\%$ at the dSCGF($3$) level.
The intrinsic deformations of the oblate ground state and of the prolate solution are remarkably stable with respect to both the basis size and the many-body truncation order. The excitation energy of the latter is consistent with the observed $0^+_3$ level, supporting its interpretation as a prolate shape isomer.
 
Several developments will naturally follow this work. 
The short-term objective is to use the presently reported framework to extend the reach of ab initio nuclear structure calculations to (very) heavy nuclei. This will constitute a major breakthrough.  
Extending the present analysis to the spectral content of the correlated propagator, i.e.\ to one-nucleon separation energies, spectroscopic factors and, ultimately, to deformed optical potentials, constitutes a second natural direction.
Eventually, an important mid-term objective concerns the restoration of the broken rotational symmetry: while intrinsic states already provide an excellent approximation of bulk properties of $J^{\pi}=0^+$ bandheads, resolving the individual members of the rotational band requires an explicit angular-momentum projection, which remains to be formulated and implemented within SCGF theory.

\section*{Acknowledgments}
A.S.\ acknowledges the Swedish Research Council (Grants No.~2021-04507 and No.~2025-05618) and the National Academic Infrastructure for Supercomputing in Sweden (NAISS), funded by the Swedish Research Council, for providing computational resources.
This project was also provided with HPC and storage resources by GENCI at TGCC, France, thanks to Grant No. A0190513012 on the supercomputer Joliot-Curie’s ROME partition.
The authors acknowledge A.~Ekstr\"om, C.~Forss\'en and M.~Frosini for useful discussions, and C.~Barbieri for numerical benchmarks of dSCGF($3$) calculations against sSCGF($3$) ones in doubly closed-shell nuclei.

\bibliographystyle{apsrev4-2}
\bibliography{biblio}

\end{document}